\documentstyle[preprint,aps,floats]{revtex}
\ifx\nopictures Y \else \input epsf.tex \fi
\tighten
\begin{document}
\def\beq{\begin{equation}}
\def\eeq{\end{equation}}
\def\D0{D\0~}
\def\ov{\overline}
\def\ra{\rightarrow}
\def\pslash{\not{\hbox{\kern -1.5pt $p$}}}
\def\wslash{\not{\hbox{\kern -4pt $\cal W$}}}
\def\tta{{\mbox {\,$t$-${t}$-$A$}\,}}
\def\Tr{{\rm Tr}}
\def\fb{${\rm fb}^{-1}$}
\def\ETslash{\not{\hbox{\kern-4pt $E_T$}}}
\def\gino{\tilde g}

\setcounter{footnote}{1}
\renewcommand{\thefootnote}{\fnsymbol{footnote}}

\preprint{MSUHEP-80226, \\ hep-ph/yymmnn}
\title{
Signatures of the light gluino in the top quark production}
\author{ Chong Sheng Li$^a$,  P. Nadolsky$^b$, C.-P. Yuan$^b$, Hong-Yi Zhou$^c$}
\address{
$^a$\rm Department of Physics, Peking University,
Beijing 100871, China
}
\address{$^b$\rm Department of Physics and Astronomy, Michigan State University,\\
East Lansing, MI 48824, USA}
\address{
$^c$\rm Institute of Modern Physics and Department of Physics,\\ Tsinghua University,
Beijing 100084, China}

\date{\today}
\maketitle
\begin{abstract}
If a light gluino, with a mass of the order of GeV, exists in the 
minimal supersymmetric extension of the Standard Model, then 
it can contribute to the production rate of the top quark pairs at
hadron colliders via $\gino \gino \ra t \bar t$. Because the top quark
is heavy,
the masses of the superpartners of the left-handed and right-handed top 
quarks can be very different such that a parity-violating observable can
be induced in the tree level production process. We discuss the phenomenology
of this parity violatiing asymmetry 
at the CERN Large Hadron Collider.
\end{abstract}
\newpage 

\section{Introduction}

In despite of the success of the Standard Model (SM) in explaining and
predicting experimental data, it is widely believed that new physics
has to set in at some high energy scale. One of such new physics models 
is the the minimal supersymmetric extension of the Standard Model (MSSM).
Various supersymmetry (SUSY) models, such as gravity-mediated
and gauge-mediated supersymmetry breaking models \cite{gunion}, 
have been extensively considered 
in the literature to explain why the 
masses of the superparticles are not the same as those of the SM particles. 
In general, 
the masses of the superparticles are predicted to be around a few hundred
GeV or at the TeV region. There are ample studies in the literature 
to examine the detection of these non-standard particles in the
current and future experiments, including those at 
the CERN Large Hadron Collider (LHC), and at the future 
Linear Colliders.

Among the superparticles of the MSSM, some models of SUSY breaking 
predict the existence of a light gluino with the masses 
around 1 GeV or less \cite{lg}. 
If this scenario is true, then there is rich phenomenology 
predicted for the current experimental data which can be used to 
either confirm or constrain models.
In Ref.~\cite{ALEPH},  
ALEPH Collaboration used the data on the cross-sections 
of dijet production and the angular distributions 
in 4-jet production to derive the ratios of the color 
factors $C_A/C_F$ and $T_F/C_F$. Based on the obtained values, 
ALEPH excluded the existence 
of the gluinos with the mass lighter than $6.3$\,GeV at $95\%$ confidence level. 
The result was criticized by Farrar~\cite{Farrar}, who argued that 
ALEPH's analysis underestimated the theoretical uncertainties in the knowledge  
of hadronization  and resummation of large logarithms arising 
in the separation of jets from soft radiation. If these uncertainties 
are taken into account, the light gluino is excluded only at $1\sigma$ level.
This problem was further examined by
Csikor and Fodor in Ref.~\cite{Csikor97}, where they determined the color 
factors \quad of underlying gauge theory\quad  
by studying \quad the behavior of the ratios $R_\gamma=\sigma(e^+e^- 
\to jets)/\sigma(e^+e^-  \to \mu^+\mu^-)$, \qquad $R_\tau=\Gamma(\tau^- 
\to  \nu_\tau + jets)/\Gamma(\tau^- \to \nu_\tau e^-  \bar\nu_e)$, 
$R_Z=\Gamma(Z\to hadrons)/\Gamma(Z \to \mu^+\mu^-)$ in the region 
of $5$\,GeV to $M_Z$ scale.      
They concluded that the ${\cal O} (\alpha^3_s)$ analysis  of these quantities allows 
to exclude the light gluino with the mass between $3$ and $5$\,GeV at  
$93 \%$ confidence level, and with the mass less than $1.5$\,GeV at $70.8 \% $ 
confidence level. If their results are combined with the $\chi^2$-distribution
from ALEPH analysis, the exclusion confidence level is improved to $99.97\% $ and $99.89\% $, 
respectively. This conclusion is quite insensitive to the
overall error of ALEPH's results; for instance, the 
exclusion limits of the combined analysis are still above 
$95\% $ if ALEPH's systematic error is increased by
a factor of 3.
However, in order to extract the number of active
fermions from the experimental data, both methods
\cite{ALEPH} and \cite{Csikor97} rely on the state-of-art usage
of perturbative theory. None of the separate analyses can
exclude the light gluino at the
confidence level $\geq 70 \%$, and combined and complicated analysis is
needed to overcome the flaws of each separate method.  

Another significant limitation on the possible parameter space
of the models with light gluinos 
was recently imposed by the negative results of the search for
the production of charginos with the mass less than $m_W$ at 
LEP2 \cite{LEP2}. This result disfavors the models with the masses 
of all gauginos vanishing at tree level at GUT scale~\cite{lg}, in
which gluino has the mass of the order GeV, and  at the electroweak scale
at least one of the charginos is necessarily lighter than $W$-boson.
However, the LEP2 data can not rule out the models with
the other spectra of gaugino masses, for instance,
the models of gauge-mediated symmetry breaking 
where the gluino can be the only light gaugino \cite{RMN}.
As mentioned before, the analysis of \cite{ALEPH,Csikor97}
already puts strong constraints on the possibility of the light
mass of the gluino, however, due to the aforementioned theoretical
difficulties it seems that more study is needed. 


There are a few other methods discussed in the literature to
look for light gluino.
If gluino is light and hadronizes before reaching the detector,
it should be possible to observe its bound states, for example,
$R^0$-mesons, created by binding of gluon and gluino~\cite{FarR0}.
Although the region of $R^0$ masses is significantly restricted by
KTeV measurements~\cite{KTeV},  $R^0$ can still exist in the mass
region $1.4 -  2.2$ GeV~\cite{Far2}.

If the squark masses are of the order of several
hundreds GeV, the light mass of gluino can lead to the
noticeable peaks in the dijet invariant mass and angular
distributions at TEVATRON or LHC, 
arising due to the resonant production of massive squarks 
in the quark-gluino fusion~\cite{dijet}. The
already existing TEVATRON data allows to exclude the
light gluino models with the masses of the lighter squarks   
lying between 150 and 650 GeV~\cite{dijet};
it would be desirable to continue the search for
the resonant peaks at TEVATRON, as well as at LHC, where the increased
dijet production cross-section 
would allow to cover a 
larger region of squark masses.

In the pQCD theory, the existence of a light gluino would change the
running of the strong coupling, as well as the
form of the Dokshitser-Gribov-Lipatov-Altarelli-Parizi (DGLAP) equations.
Therefore, to describe the existing DIS data, it is necessary 
to account not only for the quark and gluon distribution functions 
inside the initial hadron(s), but also
for the gluino distribution which has 
different renormalization group properties.
An obvious question is whether the currently available 
hadronic data is consistent with the existence of 
light gluino. The last analysis of this type
was done in 1994 publications~\cite{Botts,Vogt94}, which
showed that the existence of the light gluino
didn't contradict DIS data available at that time.
However, those analyses did not include the more recent
data from H1, ZEUS and NMC experimental groups~\cite{H1,ZEUS,NMC} 
covering the region of lower $x$ and $Q^2$.
These new data can be crucial for testing the scenario of having a
light gluino in the supersymmetry models, because the existence of a light
gluino would imply a slower running of the parton distribution functions
from the low to high $Q^2$.

The existence of new 
types of particle interactions can be proved if  one observes 
the violation of the
symmetries of the Standard Model,
for instance, the significant violation of the discrete symmetry
with respect to space reflections ($P$-parity) in strong interactions. 
Experimental search for parity-violating effects could be performed
relatively easy in the processes with $t$-quarks in the final state,
due to the possibility to trace the polarization of the tops 
decaying  through the channel $t\to W^+ + b$. 
Therefore, in this work we would like to concentrate 
on the production of top quark pairs.
For the top quark pairs produced at hadron colliders,
the SM allows the production processes $q\bar q,GG \to G \to t \bar t$
to violate $P$-parity in the next-to-leading orders due to the 
presence of $W$ and $Z$ bosons in loop
diagrams. However, this effect is shown to be negligible \cite{SMP}.
On the other hand, for certain choices of SUSY parameters in the MSSM, 
it is possible to
obtain a large difference  between the masses of right-stop and left-stop,
which in principle can lead to some noticeable asymmetries
in the production of right- and left-handed top quarks. These asymmetries
arise either in the next-to-leading order of the SUSY QCD process 
$q\bar q,GG \to G \to t \bar t$, 
or at the tree level of the SUSY QCD process
$\gino \gino \to t \bar t$.
The asymmetries of the first type were studied earlier in~\cite{Loop}.
It was  shown  that at TEVATRON the difference  in the cross-sections 
of right- and left-handed $t$-quark production can be of the 
order $2-3\%$ provided
the right-stop is light. This conclusion holds 
for a wide range of gluino masses.  
As it will be shown below,
the asymmetries of the second type can only be noticeable if
gluinos are light, and the parton density of gluinos 
in the nucleon is comparable with 
that of the sea quarks.

The primary goal of this article is to present the leading order (LO) 
study of the second scenario, and to evaluate  
the impact of the small 
mass of gluino on the production
of $t$-quarks at the CERN Large Hadron Collider (LHC). 
In the first part of this study we obtained the 
LO distributions of the partons in the nucleon with the account for the
possible non-zero contents of light gluinos.
For this purpose we modified the fitting program used previously
to obtain CTEQ4L parton distributions~\cite{Lai}.  Since our study is the
leading order calculation, we considered it sufficient not to perform 
the complete
NLO analysis of parton distributions, contrary to what was done
in~\cite{Botts,Vogt94}. 

In the  course of the study, it was a surprise for us to find that
the account for the new hadronic data from H1, ZEUS and NMC 
groups~\cite{H1,ZEUS,NMC}, which was not available at the time
of the previous studies \cite{Botts,Vogt94}, tends to increase
the overall $\chi^2$ of the fit after the inclusion of a light
gluino. The reason for this is that these new data cover the 
region of lower $x$ and $Q^2$, thus making the analysis more
sensitive to the slower running of the parton distributions in
the SUSY QCD theory with a light gluino.
Nonetheless, we would like to be  extremely cautious about 
this observation and refrain from
any final conclusions about the consistency of 
the current experimental data and 
the SUSY QCD theory with a light gluino before
more thorough next-to-leading order global analysis of hadronic
data 
is made. Instead, we would like to concentrate on the primary goal
of this paper, namely, on the calculations of the top quark
production asymmetries at the LHC. For this process, the 
Bjorken $x$ of the initial state partons are 
allowed to lie in the range 
\begin{equation}
\frac{4 m^2_t}{s} = 6.25\cdot 10^{-4} \leq x_{1,2} \leq 1, 
\end{equation}
where
the parton distributions are less dependent on the low $x$
data. We therefore 
expect the results of this work
to be  stable with respect to the possible changes
in the parton distributions, and that these changes will not
introduce an uncertainty more important than those coming
from the other sources (e.g. next-to-leading order corrections).

After obtaining the parton distribution functions (PDFs) 
in the SUSY QCD theory with a light gluino, we calculate the
degree of parity violation in the $t \bar t$ pairs produced via the 
LO $\gino \gino \to t\bar t$.
Thus, the paper consists of 3 main sections:  the
description of the parton distribution functions for the 
SUSY QCD theory with a light gluino, the calculation of the
cross-sections for the process $pp (\gino \gino) \to t_{L,R}\bar t$, 
and the numeric analysis of the asymmetries in the left- and 
right-handed top quark production. Finally, the conclusion
summarizes the obtained results.

\section{Parton distributions}
We start the construction of parton distributions by assuming 
that the only superparticle, actively present in the
nucleons at the energies of the supercolliders, is gluino, with its mass
much smaller than the typical scales of $t\bar t$ production 
( less than  $1.5$ GeV 
compared to $m_t \approx 175$ GeV). 
For the purpose of our calculation, 
we incorporate the gluino sector into 
the PDF evolution package, 
used recently  to build the set of CTEQ4 unpolarized
parton distributions \cite{Lai}. 
In order to simplify the modifications
in the fitting program, we used the approach close to the one 
adopted by the authors of GRV distributions \cite{Vogt94}.
The input scale $Q_0$ for the parton distributions was chosen to be
lower than in CTEQ4L and equal to the mass of gluino (assumed to
be $m_{\gino}=0.5$ GeV in this study, unless stated otherwise). 
At this scale, the only input distributions are 
of gluons and lighter ($u,\ d,\ s$) quarks, while
the non-zero PDFs of  gluinos and heavy quarks  are 
radiatively generated at scales above their mass thresholds.

In the presence of light gluino, two aspects of the leading-order 
evolution of
parton distributions are different from that in the standard QCD. First, the
one-loop $\beta$-function,  
determining the running of the strong coupling $\alpha_s$,
\begin{equation}
 \alpha_s(Q^2) = \frac{4\pi}{\beta_0 \ln (Q^2/\Lambda^2)},
\end{equation}
 now has the form
\begin{equation}
\beta_0 = 11-{2\over 3}n_f-2n_{\tilde g}, 
\end{equation}
where $n_f$ and $n_{\gino}$ are the number of active quark flavors 
and gluinos,
respectively.
In our analysis,
we  use $ \alpha_s (M_Z) =0.118$ and $ \Lambda =7.65 \ MeV $
for 5 flavors. The matching 
of $\alpha_s$ between 4 and 5, and 5 and 6 flavors 
takes place at $Q=5.0$ GeV and $ Q =175$ GeV  
respectively, which are defined as the bottom and top quark masses.

Second, the leading order DGLAP equations
now should account for the splittings $\gino \to
q\bar{\tilde q}$, $\gino \to g \gino$ and $g \to \gino \gino$, 
so that the singlet equation takes the form of
\begin{eqnarray}
 \frac{d}{dt}\left( \begin{array}{c}
q_S (x,Q^2)\\ G(x,Q^2))\\ \gino (x,Q^2) )\end{array}\right)= \frac{\alpha
_{S}(Q^2)}{2\pi }\,\int^{1}_{x} \left(\begin{array}{ccc}
P_{qq}\bigl(x/y\bigr)  & P_{qG}\bigl(x/y \bigr) &
 P_{q\gino} \bigl(x/y \bigr)\\ 
P_{Gq}\bigl(x/y \bigr) & P_{GG}\bigl(x/y\bigr) &
 P_{G\gino }\bigl(x/y\bigr)  \\
P_{\gino q}\bigl(x/y \bigr) & P_{\gino G}\bigl(x/y\bigr) &
 P_{\gino \gino }\bigl(x/y
\bigr)
\end{array}\right) \nonumber \\ 
\times \left( \begin{array}{c} 
q_S (y,Q^2)\\  G(y,Q^2) \\ \gino (y,Q^2) \end{array} \right)\,
\frac{dy}{y}. \label{SAP}
\end{eqnarray}
The splitting functions used in (\ref{SAP}) can be found, e.g.,
 in \cite{Ant}.

With the help of the upgraded evolution package we performed the 
fit of the experimental data, 
closely following
the procedure of the construction of CTEQ4L PDF set, as described in 
\cite{Lai}. 
However, for simplicity, the fit didn't use the jet data,
and the value of the strong coupling was fixed to be 
equal to the world-average value $\alpha_s (M_Z)= 0.118$.
As a result, we obtained the set of parton distributions SUSYL, 
which was used
throughout the rest of the paper.

\section{Calculation of the matrix elements}
At the LHC, the $t \bar t$ pairs 
will be dominantly produced from the standard QCD processes of
$q\bar q$ and $GG$ interactions, as shown in Figs.~1a-d. All of these
diagrams preserve $P$-parity.
In the SUSY QCD theory,  
if the mass of gluino is small ($m_{\gino} \sim 1$ GeV), we
will also expect a noticeable contribution due to the annihilation of
gluinos, described by the three diagrams of Figs.~1e-g. 
The $s$-channel diagram of Fig.~1e  is equivalent, up
to a color factor, to the analogous  
$q\bar q$ diagram of Fig.~1a and
doesn't break parity; however, the parity symmetry is broken in the $t$-
and $u$-channels  due to the mechanism of squark mass mixing which
is briefly described below.

In MSSM the left-squark and the right-squark, 
superpartners of the the left- and right-handed quarks, 
do not have definite mass 
but instead are a mixture of two mass eigenstates.  
These mass eigenstates $\tilde{q}_1$ and $\tilde{q}_2$ are
related to the current eigenstates $\tilde{q}_L$ and $\tilde{q}_R$ by
\begin{equation}
\tilde{q} _1 = \tilde{q} _L \cos \theta _q + \tilde{q} _R \sin \theta
_q,~~~\tilde{q} _2 = -\tilde{q} _L \sin \theta _q + \tilde{q} _R \cos
\theta _q.
\end{equation}  
Due to this, MSSM in general allows to have nonzero asymmetries in $ q
\bar q$-pair production, defined by
\begin{equation}
 A_q=\frac{\sigma (pp\to q_L \bar q)-\sigma (pp\to q_R \bar q)}
 {\sigma (pp\to q_{L}  \bar q)+\sigma (pp\to q_R \bar q)},\label{A}
\end{equation}
where $\sigma$ denotes the cross-section of the 
$t\bar t$-pair production, integrated over the relevant part of the
phase space to be discussed below.
The best chance to observe a non-zero $A_q$ is provided by the  
$t \bar t$ production process, where the mixing between the squarks
is the largest due to the 
large mass of the top quark. In the following, we ignore the mass 
mixing for the lighter 5 quarks.

In MSSM, the squark-quark-gluino interaction
Lagrangian is given by\\
\begin{eqnarray}
 L_{\tilde{g} \tilde{q} \bar{q}} & = & -g_s T^a_{jk}\bar{q}_k[(a_1-b_1
\gamma_5)\tilde{q}_{1j}+(a_2-b_2\gamma_5)\tilde{q}_{2j}]\tilde{g}_a+h.c.\;,
\label{L}
\end{eqnarray}          
where $g_s$ is the strong coupling constant, $T^a$
are $SU(3)_C$ generators and $a_1,~b_1,~a_2,~b_2$ are given by
\begin{displaymath}
  a_1=\frac{1}{\sqrt{2}}(\cos\theta _q-\sin\theta_q)=-b_2,
\end{displaymath}
\begin{equation}
 b_1=-\frac{1}{\sqrt{2}}(\cos\theta _q+\sin\theta_q)=a_2. \label{ab}
\end{equation}  

The mixing angle $\theta_t$ and the masses $m_{\tilde{t}_1}$,
$m_{\tilde{t}_2}$ can be calculated by diagonalizing the following mass
matrix:
\begin{equation}
\label{eqnum2}
         \left(
	   \begin{array} {ll}
	     M^2_{\tilde{t}_L} & m_tm_{LR} \\
	     m_tm_{LR} & M^2_{\tilde{t}_R}
	   \end{array}
         \right),   
\end{equation}
where $M^2_{\tilde t_{L,R}}$ and $m_{LR}$ are the parameters of the
soft-breaking terms in the MSSM.

From Eq.(\ref{eqnum2}), we can derive the expressions for
$m^2_{\tilde{t}_{1,2}}$ and $\theta_t$ :
\begin{eqnarray}  
& & m^2_{\tilde{t}_{1,2}}=\frac{1}{2}\left[
M^2_{\tilde{t}_{L}}+M^2_{\tilde{t}_{R}}\mp\sqrt{(M^2_{\tilde{t}_{L}}
-M^2_{\tilde{t}_{R}})^2+4m_t^2m_{LR}^2}\right], \label{m12} \\
& &\tan\theta_t=\frac{m^2_{\tilde{t}_1}-M^2_{\tilde{t}_{L}}}{m_tm_{LR}}.
\label{tht}
\end{eqnarray}
Inversely,
\begin{equation}
M^2_{\tilde{t}_{R,L}}=\frac{1}{2}\left[
m^2_{\tilde{t}_{1}}+m^2_{\tilde{t}_{2}}\mp\sqrt{(m^2_{\tilde{t}_{2}}
-m^2_{\tilde{t}_{1}})^2-4m_t^2m_{LR}^2}\right]. \label{mLR} 
\end{equation}

The asymmetry $ A_t$ depends on the angle of mixing $ \theta_t$
in the following manner. The denominator of $A_t$ is dominated
by the large contributions from the quark and gluon channels (Figs.1a-d)
and therefore shows little  dependence on the masses of stops.
The numerator of the asymmetry depends both on the splitting and the
mixing of the squark masses. 
Since left/right-handed quarks couple 
only to left/right squarks, in the case of  no mass mixing
($m_{LR}=0$)
the asymmetry is completely determined by the difference of masses 
$M_{\tilde{q}_L}- M_{\tilde{q}_R}$.  In this limit 
$\theta_t \approx -\pi/2$, provided that 
$M_{\tilde{t}_{R}} < M_{\tilde{t}_{L}}$.

For  fixed mass eigenvalues $ m_{\tilde{t}_{1,2}}$,
the relationship (\ref{mLR}) for the mass parameters 
$ M_{\tilde{t}_{L,R}}$
puts the upper bound on $ m_{LR}$:
\begin{equation}
m_{LR} \leq  {m^2_{\tilde{t}_2}-m^2_{\tilde{t}_1}  \over 2 m_t}.
\end{equation}
For  largest $ m_{LR}$, 
\begin{equation}
 \theta_t =  - {\pi \over 4} 
\end{equation}
and 
\begin{equation}
 M_{\tilde{t}_L} = M_{\tilde{t}_R}.
\end{equation}
In this limit, 
stop mass eigenstates  
have the maximal mixing between the left- and right-stops,
so that the asymmetry $ A_t$ becomes zero. 
Thus, for fixed mass eigenstates, the asymmetries are expected 
to decrease with the growth of $m_{LR}$. 

In gravity-mediated supersymmetry breaking models (mSUGRA), 
the masses of left- and right-stops satisfy the relations
\begin{eqnarray}
\nonumber\\
& & M^2_{\tilde{t}_L}=m^2_{\tilde{t}_L} + m^2_t+(\frac{1}{2}
  -\frac{2}{3}\sin^2\theta_W)\cos(2\beta)m_Z^2,\nonumber\\
& &  M^2_{\tilde{t}_R}=m^2_{\tilde{t}_R} + m^2_t
  +\frac{2}{3}\sin^2\theta_W\cos(2\beta)m_Z^2,\nonumber\\
& & m_{LR}=-\mu\cot\beta+\lambda_t,     \label{M}
\end{eqnarray}   
where $m^2_{\tilde{t}_L},~m^2_{\tilde{t}_R}$ are the soft SUSY-breaking
mass terms of left- and right-stops,  $\mu$ is the coefficient
of the  $H_1$-$H_2$ mixing term in the superpotential, $\lambda_t$
is the parameter describing the strength of soft SUSY-breaking trilinear
scalar interaction $\tilde{t}_L\tilde{t}_R H_2$,
$\tan\beta=v_u/v_d$ is the ratio of the vacuum expectation values
of the two Higgs doublets.
In the minimal supergravity models, the soft SUSY breaking
parameters $m^2_{{\tilde q}_L}$ and  
$m^2_{{\tilde q}_R}$ are  equal to each other, so that the mass 
splitting $M^2_{{\tilde q}_L}- M^2_{{\tilde q}_R}$ is 
small, and of the same order of
magnitude for all quark flavors. 
In this case, it is hard to expect 
observable asymmetries. 
On the other hand, in the  general MSSM 
right- and left-squark masses $M_{\tilde t_{L,R}}$ are considered to be independent
parameters, in which case there is no theoretical limitations on 
the splitting of stop masses.  
In the following, the second point of 
view is accepted, so that  $M_{{\tilde t}_R}$ is assumed to be 
of the order 90-175 GeV,
while $M_{{\tilde t}_L}$ is varied between 150 and 1000 GeV.

The cross-sections entering the asymmetry (\ref{A}) are calculated in a
usual way by convolution of the squared and 
spin- and color-averaged hard scattering matrix elements 
$ |{\cal M}_{k_1 k_2}|^2_{L,R}$ with
the appropriate parton distributions $f_{i}(x)$:
\begin{equation}
 \sigma({pp\to t_{L,R} \bar t}) =
\frac{\beta}{32 \pi \hat s}\int_{-1}^{1}d\cos 
{\theta\int{dx_1 dx_2\sum_{i_1,i_2}{f_{i_1}(x_1)f_{i_2}(x_2)
|{\cal M}^{i_1 i_2}(\hat s, \hat t,\hat u)|^2_{L,R}}}},  \label{sig}
\end{equation}
where $ \hat s, \hat t,\hat u$ are the parton Mandelstam variables,  
$ \beta \equiv \sqrt{1-{4m_t^2}/{\hat{s}}}$
and the particle momenta  for the partons $q_{i_{1,2}}$ in the initial state
are defined as $q_{i_1}(p_1)  + q_{i_2}(p_2) 
\to t(p_3) + \bar t (p_4)$.
The numerator of the asymmetry (\ref{A}) is determined solely by 
the diagrams Figs.~1f,g (containing stops), which give the following 
matrix elements for the production of left-handed top:
\begin{eqnarray}
  &&\Biggl( {\cal M}_t {\cal M}_u^{\dagger} + {\cal M}_u  {\cal M}_t^{\dagger} \Biggr)_L =
  \sum_{i,j} \frac{a^2_i a^2_j}{(\hat t - m^2_{\tilde t_i})
   (\hat u - m^2_{\tilde t_j})} \nonumber \\ 
&& \times \Biggl(  4 m^2_t {\hat s} (1-C_i C_j)\Bigl( (1-C_i
C_j)+(C_i-C_j)\cos\theta \Bigr) \nonumber \\
&&  -2 (m^2_t - {\hat t})(m^2_t -{\hat u})(1-C_{i}^2)(1-C_{j}^2) \Biggr), \label{tu}
\end{eqnarray}
\begin{equation} 
 | {\cal M}_t|^2_L =\sum_{i,j}\frac{a^2_i a^2_j}
  {2(\hat t - m^2_{\tilde t_i})(\hat t - m^2_{\tilde t_j})}
(m^2_t -\hat t) \hat s (1+C_i C_j) \left( A+B \cos \theta
\right), \label{t} 
\end{equation}
\begin{equation}
 | {\cal M}_u|^2_L
  =\sum_{i,j}\frac{a^2_i a^2_j} {2 (\hat u - m^2_{\tilde t_i})(\hat u -
  m^2_{\tilde t_j})} (m^2_t -\hat u) \hat s (1+C_i C_j)
\left(A -B \cos \theta \right). \label{u}
\end{equation}
In these formulas 
\begin{eqnarray}
&& C_i\equiv {b_i \over a_i},\quad i=1,2,\\
&& A \equiv (1+C_i)(1+C_j)(1-\beta) + (1-C_i)(1-C_j)(1+\beta),\\
&& B \equiv (1+C_i)(1+C_j)(1-\beta) - (1-C_i)(1-C_j)(1+\beta),
\end{eqnarray}
the summation ($i,j$) goes over the two stop masses.
The squared matrix element  
$|{\cal M}^{\gino \gino}|^2_{L}$
entering (\ref{sig}) can be written in terms of (\ref{tu}-\ref{u}) as
\begin{equation}
|{\cal M}^{\gino \gino}|^2_{L} = \frac{1}{256}\biggl(
\frac{16}{3}\Bigl(|{\cal M}_t|^2 + |{\cal M}_u|^2 \Bigr) + 
\frac{2}{3}\Bigl( {\cal M}_t {\cal M}_u^{\dagger} + {\cal M}_u  {\cal M}_t^{\dagger}\Bigr)
\biggr)_L.
\end{equation}
The matrix elements for the production of right-handed
 top are obtained by the
substitution
\begin{equation}
  C_{i,j} \to -C_{i,j} .
\end{equation}

If (\ref{tu}-\ref{u}) are combined with the explicit 
formulas (\ref{ab}) for $a_i,\ b_i$,
it is possible to get the following expression for 
the difference of the matrix elements for producing 
the left- and right-handed top quarks in the $t \bar t$ pairs:
\begin{eqnarray}
&& | {\cal M}^{\gino\gino}|^2_L - 
| {\cal M}^{\gino \gino}|^2_R = 4\cos{2 \theta_t}\biggl\{(X_{11}-X_{22})
(\beta- \cos\theta)\nonumber \\
&& +(Y_{21}-Y_{12})\cos\theta +(Z_{11}-Z_{22})(\beta+\cos\theta)\biggr\},
\label{c2t}
\end{eqnarray}
where 
\begin{eqnarray}
X_{ij} &\equiv& 
\frac{(m^2_{t}-\hat t)\hat s}{96 (\hat t - 
m^2_{\tilde t_i})(\hat t - m^2_{\tilde t_j})},\\
Y_{ij} &\equiv& 
\frac{m^2_{t}\hat s}{192 (\hat t - m^2_{\tilde t_i})
(\hat u - m^2_{\tilde t_j})},\\
Z_{ij} &\equiv& 
\frac{(m^2_{t}-\hat u)\hat s}{96 (\hat u - m^2_{\tilde t_i})
(\hat u - m^2_{\tilde t_j})}.
\end{eqnarray}
Equation (\ref{c2t}) depends on the mass mixing angle $\theta_t$
only through the common factor $\cos{2\theta_t}$. This  proves
the argument given before that for fixed $m_{\tilde{t}_{1,2}}$ the asymmetry 
should be the largest at $m_{LR}=0$ and $\theta_t = -\pi/2$.

The diagrams in Figs.~1a-e do not violate the parity and need to be included 
only in the denominator of the asymmetry (\ref{A}). 
The matrix elements for the pure QCD
processes (in Figs.~1a-d) are well-known, while the $s$-channel $
\gino\gino$ diagram (in Fig.~1e) differs from the analogous $q\bar 
q$ one only by a color factor:  
\begin{eqnarray}
&& | {\cal M}^{q\bar q}|^2={4\over 9}\frac{ 
(m^2_t-\hat t)^2+(m^2_t-\hat u)^2+
2 m^2_t  {\hat s} }{ \hat s^2},\label {qq}\\
&&   | {\cal M}^{GG}|^2 ={1\over 16}\Bigl( \frac{
(m^2_t-\hat t)(m^2_t-\hat u)}{ 12 \hat s^2}+
{8\over 3}\frac{(m^2_t-\hat t)(m^2_t-\hat u)- 2 m^2_t (m^2_t +\hat t) 
}{ (m^2_t-\hat t)^2} \nonumber \\
 && + \frac{8}{3}\frac{(m^2_t-\hat t) (m^2_t-\hat u)-
         2 m^2_t (m^2_t +\hat u)}{ (m^2_t-\hat u)^2}-
{2\over 3} \frac{m^2_t  (\hat s-4 m^2_t)}
{ (m^2_t-\hat t)(m^2_t-\hat u)}\nonumber \\
  &&    -6 \frac{(m^2_t-\hat u)(m^2_t-\hat t)+
m^2_t (\hat u-\hat t)}{ \hat s (
m^2_t-\hat t)}\nonumber \\
&& -6 \frac{(m^2_t-\hat u)(m^2_t-\hat t)-m^2_t  (\hat u-\hat t)}{ 
\hat s (m^2_t-\hat u)}\Bigr),\\
&& | {\cal M}_s^{  \gino\gino  }|^2 = {27\over 32}  | 
{\cal M}^{q\bar q}|^2.\label{s}
\end{eqnarray}
In the above, the spin and color factors in both the final 
and the initial states are all properly summed and averaged.

One can also obtain the total parton cross-sections by the integration of 
(\ref{tu}-\ref{u},\ref{qq}-\ref{s}) over the scattering angle $ \theta$. 
For the
$t$ and $u$-channels we define 
\begin{eqnarray}
& & C \equiv 2(1-\beta^2)(1-C_i C_j)(C_i-C_j)\\
& & D \equiv \beta^2(1-C_i^{2})(1-C_j^{2})\\
& & E  \equiv 2(1-\beta^2)(1-C_iC_j)^2-(1-C_i^2)(1-C_j^2), \\
& & v_i (\hat s, \beta) \equiv \frac{2m_{\tilde t_i}^2+\hat{s}+
\beta\hat{s}-2m_t^2}
        {2m_{\tilde t_i}^2+\hat{s}-\beta\hat{s}-2m_t^2}.
\end{eqnarray}
Then
\begin{eqnarray}
& & [\sigma^{\gino\gino}_{tu}]_L=\sum\limits_{i,j}\frac{g_s^4a_i^2a_j^2}
{24576\pi}\big[8
(1+C_iC_j)f_1(m_{\tilde t_i}^2,m_{\tilde t_j}^2)+f_2(m_{\tilde 
t_i}^2,m_{\tilde t_j}^2)\big], \\
&& f_1(m_{\tilde t_i}^2,m_{\tilde t_j}^2) = \frac{1}{\beta(m_{\tilde 
t_i}^2-m_{\tilde t_j}^2)}\bigg(
-\frac{8B}{\hat{s}}\beta(m_{\tilde t_i}^2-m_{\tilde t_j}^2) \nonumber \\
& &+4(m_{\tilde t_i}^2-m_t^2)(2Bm_{\tilde t_i}^2+
B\hat{s}+A\beta\hat{s}-2Bm_t^2)/\hat{s}^2\ln 
v_i(\hat s, \beta) \nonumber\\
& &-4(m_{\tilde t_j}^2-m_t^2)(2Bm_{\tilde t_j}^2+
B\hat{s}+A\beta\hat{s}-2Bm_t^2)/\hat{s}^2
\ln v_j(\hat s, \beta)\bigg),
\end{eqnarray}
\begin{eqnarray}
&& f_2(m_{\tilde t_i}^2, m_{\tilde t_j}^2) = \frac{1}{\beta^2}\bigg\{
-\frac{8D}{\hat{s}}\beta+\bigg(\bigg[
2E\beta^2\hat{s}^2+(4m_{\tilde t_i}^2+2\hat{s}-4m_t^2)
\nonumber \\
& & \times(2Dm_{\tilde t_i}^2+D\hat{s}+C\beta\hat{s}-2Dm_t^2)\bigg]
\ln v_i(\hat s, \beta)\nonumber \\
& & +\bigg[2E\beta^2\hat{s}^2+(4m_{\tilde t_j}^2+2\hat{s}-4m_t^2)\\
& & \times(2Dm_{\tilde t_j}^2+D\hat{s}-C\beta\hat{s}-2Dm_t^2)\bigg]
\ln v_j(\hat s, \beta) \bigg)\nonumber \\
& & \times\frac{1}{\hat{s}^2(m_{\tilde t_i}^2+
m_{\tilde t_j}^2+\hat{s}-2m_t^2)}
\bigg\}.
\end{eqnarray}
When $m_{\tilde t_i}=m_{\tilde t_j}$, we have
\begin{eqnarray}
&& f_1(m_{\tilde t_i}^2=m_{\tilde t_j}^2) = \frac{1}{\beta}\bigg\{
-\frac{8B}{\hat{s}}\beta 
+4(4Bm_{\tilde t_i}^2+B\hat{s}+A\beta\hat{s}-4Bm_t^2)/\hat{s}^2
\ln v_i (\hat s, \beta)\nonumber \\
& &+8(m_{\tilde t_i}^2-m_t^2)(2Bm_{\tilde t_i}^2+
B\hat{s}+A\beta\hat{s}-2Bm_t^2)
  \nonumber \\
& & \times (\frac{1}{2m_{\tilde t_i}^2+\hat{s}+
\beta\hat{s}-2m_t^2}-\frac{1}
{2m_{\tilde t_i}^2+\hat{s}-\beta\hat{s}-2m_t^2})/\hat{s}^2
\bigg\}.
\end{eqnarray}
Again, $[\sigma^{\gino\gino}_{tu}]_R$   can be obtained by the substitution 
\begin{equation}
 C_i\to -C_i,~~~C_j\to -C_j~.
\end{equation}

The cross-sections of the other sub-processes,
 corresponding to (\ref{qq}-\ref{s}), are given by
\begin{eqnarray}
& &\sigma^{q\bar q}= \frac{g_s^4}{108\pi\hat{s}}
\beta(2+\rho)\\
& &\sigma^{GG}= \frac{g_s^4}{48\pi\hat{s}}
\bigg((1+\rho+\frac{\rho^2}{16})\ln\frac{1+\beta}{1-\beta}-
\beta(\frac{7}{4}
+\frac{31}{16}\rho)\bigg),\\
&& \sigma^{\tilde{g}\tilde{g}}_s = \frac{g_s^4}{128\pi\hat{s}}
\beta(2+\rho),
\end{eqnarray}
where $\rho \equiv 4m^2_t/\hat s$.

\section{Numeric results}
To estimate the largest possible asymmetries, 
we varied the squark mass eigenvalues $m_{\tilde{t}_{1,2}}$ with $m_{LR}$ 
set to be zero (see the discussion in the previous Section).  
No assumption was made about any model-specific relationships between
the values of the mass parameters $M_{\tilde t_{L,R}}$ and $m_{LR}$
(c.f. eq. (\ref{M})).

If $m_{LR}=0$, the left/right-handed quarks couple independently 
to the left/right-stops. Correspondingly, for $m_{\tilde t_{1}} 
\neq m_{\tilde t_{2}}$, the production rates of the left- and right-handed
tops will be different. The asymmetry $A_t$ is expected to grow when 
the mass splitting $m_{\tilde t_{1}} 
\neq m_{\tilde t_{2}}$ increases. 
In this work, the asymmetries were calculated for 
 $m_{\tilde t_1} = 90$ GeV 
(which is consistent with the current LEP2  data \cite{LEP2}), 
$m_{\tilde t_1} = m_t = 175$ GeV and various  values of $m_{\tilde t_2}$.
Two values of factorization scale $\mu= m_t$ and $2 m_t$
were used. Various sets of masses will be further denoted 
as $(m_{\tilde{t}_1}, m_{\tilde{t}_2}, m_{LR})$, with  numerical values in 
GeV. As before, the gluino mass is assumed to be equal to 
0.5\,GeV.  

In the SM, both $t$ and $\bar t$ decay into $b$ ($\bar b$) 
and $W^\pm$ with an almost unit probability, with a subsequent decay of
the $W$-bosons into 2 jets or 2 leptons. 
In the MSSM, when both the gluino and the stop are light, the top quark
can also decay via $t \to \gino {\tilde t_1}$, so that the branching ratio
of $t \to W^+ + b$ decreases. Assuming that all the other supersymmetric
particles are heavier than the top quark, 
and $\theta_{t}= - \pi/2$, the branching ratio for
$t \to W^+ + b$ is equal to 0.29 and 1 for $m_{\tilde t_1} = 90$\,GeV
and $175$\,GeV, respectively. 
The CDF collaboration has measured 
the branching ratio of $t \ra W^+ + b$ to be 
$0.87{^{+0.13}_{-0.30}}{^{+0.13}_{-0.11}}~$~\cite{br}.
Hence, the chosen sets of the values for $m_{\tilde t_1}$ and
$m_{\gino}$ are still allowed by data within 95\% c.l.

It is convenient to study the asymmetry $A_t$ using the semileptonic 
modes of decay, with 
$t\to b {\it l}^+ \nu_{\it l}$ (${\it  l}=e, \mu$) and $\bar t\to 
\bar b q \bar q$ (or vice versa),
which have a branching ratio of about  
0.086 and 24/81 for $m_{\tilde t_1} = 90$\,GeV
and $175$\,GeV, respectively. 
In the following we assume that it will be possible to reconstruct
the kinematics of $t\bar t$-pair from the momenta of the decay products
by requiring the transverse momenta  
$p_T^{jets} \geq 30$\,GeV, $p_T^{leptons} \geq 20$\,GeV, 
$\ETslash \geq 20$\,GeV,
the rapidities of the jets and leptons $|y| <2.0$, and 
the jet cone separation $\Delta R > 0.4$~\cite{ATLAS}. 
We also assume that it 
will be necessary to tag one $b$-quark with an  efficiency $C_b=50 
\%$. We estimate the statistical error in the measurement of the 
asymmetry by
\begin{equation}
\delta A_t = \frac{1}{\sqrt {{\cal L}TC_b (\sigma^{tot}_L+\sigma^{tot}_R)}},
\label{dA}
\end{equation}
where we assume the observation time $T=1\ year$ and the luminosity
${\cal L}= 100 \ fb^{-1}/year$, corresponding to the second run of the LHC.

The  imposed selection cuts and branching ratio significantly reduce the 
total cross-section of $t\bar t$ production, 
typically from around $340 \ pb$ down to 
3.5 and 12 pb for $m_{\tilde t_1} = 90$\,GeV
and $175$\,GeV, respectively. 
As an example,
Fig.~2 shows various differential  
cross-sections including the $GG$, $q\bar q$ and $\gino\gino$ subprocesses 
for the squark masses $(90, 1000, 0)$ and $\mu = 2 m_t$, obtained with the 
kinematic cuts and branching ratios applied. 
 It can be readily seen that the dominant part of the 
$t\bar t$ pairs is produced due to the gluon-gluon subprocess, which 
contributes around $71 \%$ of the total rate. 
The quark-antiquark and gluino-gluino shares are $22\%$ and $7 \% $, 
respectively.  
The gluino contribution is  
comparable with the conventional QCD uncertainties in the knowledge 
of the total rate (about 5\% to 10\%), however, the
presence of the light gluino will change the shape of the cross-section 
distributions. 
Thus, in principle there is a  possibility to detect 
the light gluino by carefully fitting the event rate distributions
and comparing them with the predictions of perturbative 
QCD.

Fig.~3 shows the sum of the differential
cross-sections of left- and right-handed top quarks production 
$d\sigma_L/dM_{t\bar t} + d\sigma_R/dM_{t\bar t}$, and their difference
$d\sigma_L/dM_{t\bar t} - d\sigma_R/dM_{t\bar t}$ (scaled by a factor of 100),
as functions of the invariant mass of the $t\bar t$ pair $M_{t\bar t}$. 
One can see that the asymmetry is  most noticeable in the region of small and 
intermediate values of $M_{t\bar t}$. 
This is different from the behavior of 
the asymmetry produced due to the presence of superpartners in the loop corrections
\cite{Loop}. 
In that case, the asymmetry becomes significant in the region of large $M_{t\bar t}$, 
where in the case of the light gluino it 
can have the value of $2-3\%$. 
In this respect, we expect the minimal interference between the tree-level 
and loop-generated asymmetries, since the main contributions to them come
from different kinematic regions.

The dependence of the cross-sections  on two other kinematic
parameters, the transverse momentum of the $t$-quark 
and the cosine of the scattering angle in the $t\bar t$ rest frame, is 
shown in Figs. 4 and 5. 
As can be seen from Fig. 5, 
the difference $d\sigma_L/d\cos\theta - d\sigma_R/d\cos\theta $
  changes its sign around $\cos \theta\approx \pm 0.8$, 
so that one can enhance the asymmetry by separately  
considering 
the cross-sections  integrated over either large or small angles. 
 It can also be shown that  the asymmetries at  small angles 
can be further enlarged  by rejecting the events with transverse 
momenta larger than $100$ GeV$/$c. 
For the other combinations of stop masses,  $d\sigma_L/\cos\theta-
d\sigma_R/\cos\theta$ 
changes its sign at slightly lower $|\cos \theta|$, approximately $0.75-0.8$.  
We therefore present the asymmetries of the cross-sections 
integrated separately over the region 
$|\cos\theta|\leq 0.8$, or 
the region $|\cos\theta| > 0.8$ with $p_T \leq 100$ GeV$/$c.

Table 1 shows the  values of the asymmetry $A_t$ obtained after 
the integration of the rate
with the aforementioned cuts in $|\cos \theta|$ and $p_T$. As can 
be seen from the Table, for various stop masses the asymmetry ranges from 
$0.3\%$ to $1.1 \%$. 
The behavior of the asymmetries with the growth of  $m_{\tilde{t}_2}$ 
is different at large and small angles. At $|\cos\theta |\leq 0.8$ 
the asymmetry monotonously increases with the growth of  $m_{\tilde{t}_2}$, 
while at $|\cos\theta | > 0.8$ the asymmetry has a maximum around 
$m_{\tilde{t}_2} = 200$ GeV and then starts to decrease. 
At small angles ($|\cos\theta| \leq 0.8$) the asymmetry quickly decreases
with the growth of the mass of the lighter squark and becomes 
practically unnoticeable for $m_{\tilde t_{1}} \geq m_t$.

For the comparison, we also give in the same Table 
the statistical errors $\delta A_t$ from Eq. (\ref{dA}). 
These errors
are mostly determined by $GG$ and $q\bar q$  cross-sections, so
that they hardly depend on the choice of the squark masses.
For most combinations of the stop masses,
the obtained values of $A_t$ can in principle be distinguished from 
the statistical error $\delta A_t$ at a
$ 2\sigma$ level or better.
However, what can be more  important are the 
experimental systematic uncertainties related 
to the measurement of the asymmetries of the order $1\%$. In particular, 
it can be challenging to reach the necessary accuracy in  the 
reconstruction of the kinematics of the $t\bar t$-pair, 
and the determination of the top quark polarization.  
Nevertheless, the predictive 
power of this analysis can be increased if it is combined with 
the search for the signature of the light gluinos  in the other 
kinematic regions, for instance, for the loop-generated asymmetries in the 
production of the top-antitop pairs with large invariant masses.

\section{Conclusion}
In this work we propose a new method, based on the search of
the possible violations of the discrete symmetries of the Standard Model, 
to test the existence of a light gluino in the MSSM. 
This is in contrast to many other methods presented in the literature
(see the Introduction section), in which one has to assume how a light 
gluino hadronizes into hadron states to be compared with the experimental
measurement. 

We study the consequences the small mass of the gluino 
would have for the production of top quarks at the LHC via
the tree level process $\gino \gino \to t \bar t$.
We show that with a large mass splitting in the 
masses of superpartners (top-squarks) of the top quark,
the gluino-gluino fusion process
can generate the parity-violating asymmetry in the production of 
left- and right-handed $t$-quarks.
Since the SM QCD theory preserves the discrete symmetry of $P$-parity,
a small violation of such a symmetry may be observed from 
a large $t \bar t$ data sample at the LHC.

For $m_{\gino} \approx 0$ the largest values of the parity-violating
asymmetry discussed in the previous sections 
is around $0.3-1.1\%$ for various choices 
of SUSY parameters. Hence, it can in principle be observed, 
taking into the account the high rate of the top production at the 
LHC. In order to measure the asymmetry with 
a small  statistical error, the 
experiment should be preferably done during the second run of LHC
with an integrated
luminosity of $100 \ fb^{-1}/year$. 
The rate of the top quark production does not seem to be the 
major obstacle for the measurement of the parity-violating 
asymmetry. However, it demands a good understanding of the
systematic errors, better than 1\%, to reach the precision of
the measurement sufficient to test the existence of a light gluino in
$t\bar t$ pair production. 

\newpage

\section*{Acknowledgements}

We would like to thank L.~Clavelli,  L.~Dixon, G.R.~Farrar,
H.L. Lai, R.~Raja, 
T.~Rizzo, C.~Schmidt,  Z.~Sullivan and W.-K. Tung for helpful discussions.
This work was supported in part by the National Natural Science
Foundation of China, a grant from the State Commission of Science
and Technology of China, and by the U.S. NSF grant PHY-9507683.

\section*{Figure captions}
\noindent
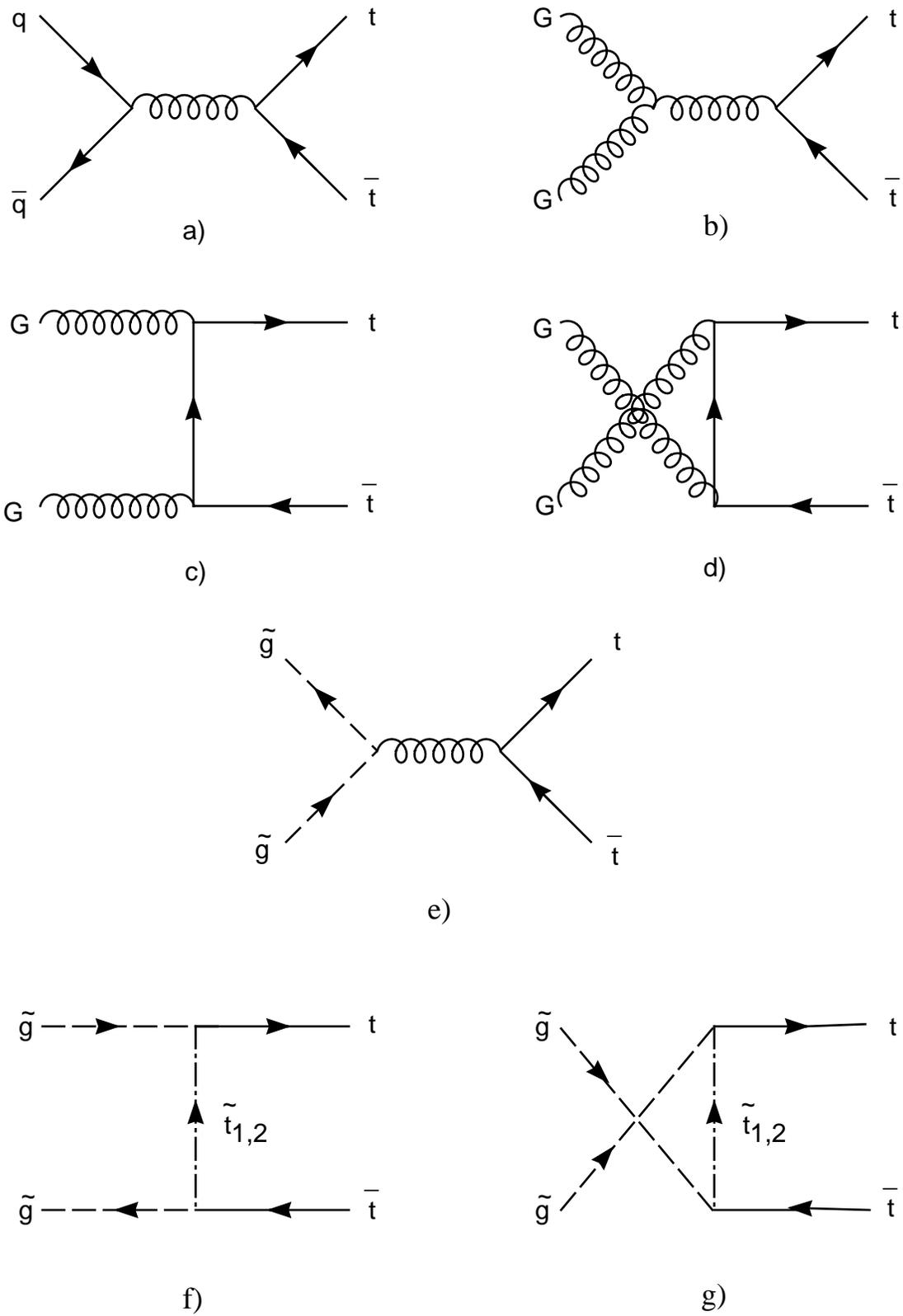
{\bf Fig. 1.} Leading order diagrams contributing 
to the production of top quarks in SUSY QCD theory.
\vspace{\baselineskip}

\noindent
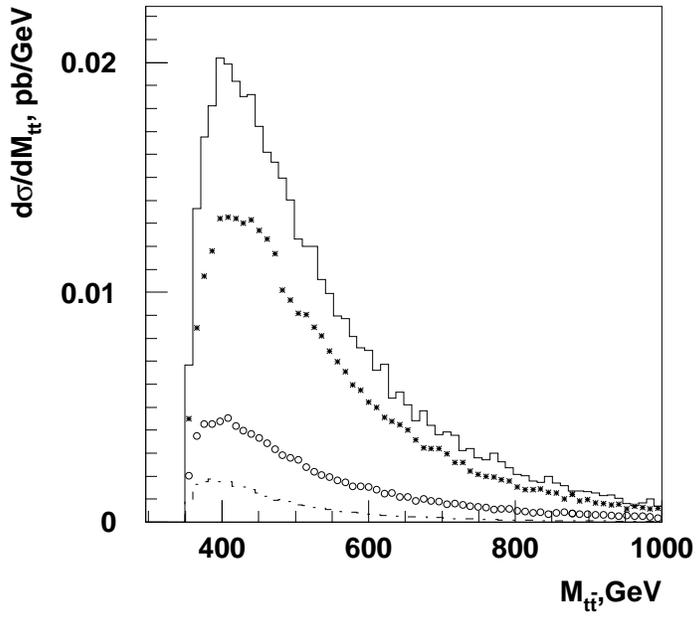
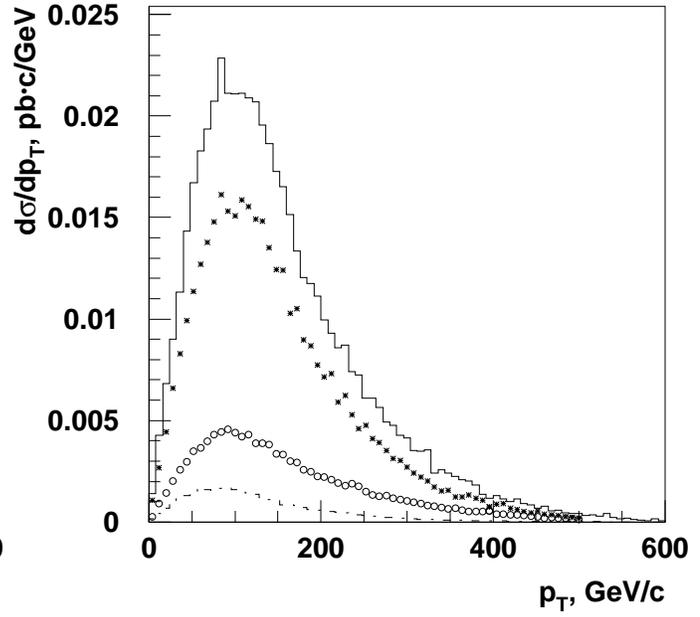
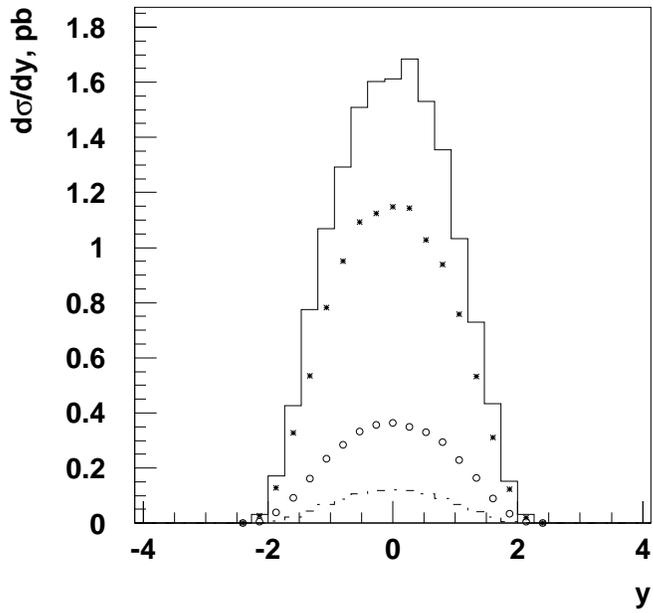
{\bf Fig. 2.} The dependence of the cross-section of $t\bar t$ 
pair production on various kinematic 
parameters: $t\bar t$ pair invariant mass $M_{t\bar t}$, $t$-quark 
transverse momentum $p_T$ and rapidity
$y$. The solid line, 
stars, circles and dashed line correspond to the full differential
cross-section 
and the contributions 
of gluon, quark and gluino subprocesses, respectively. 
The factorization scale $\mu =2 m_t$, the squark masses
are (90,1000,0).
\vspace{\baselineskip}

\noindent
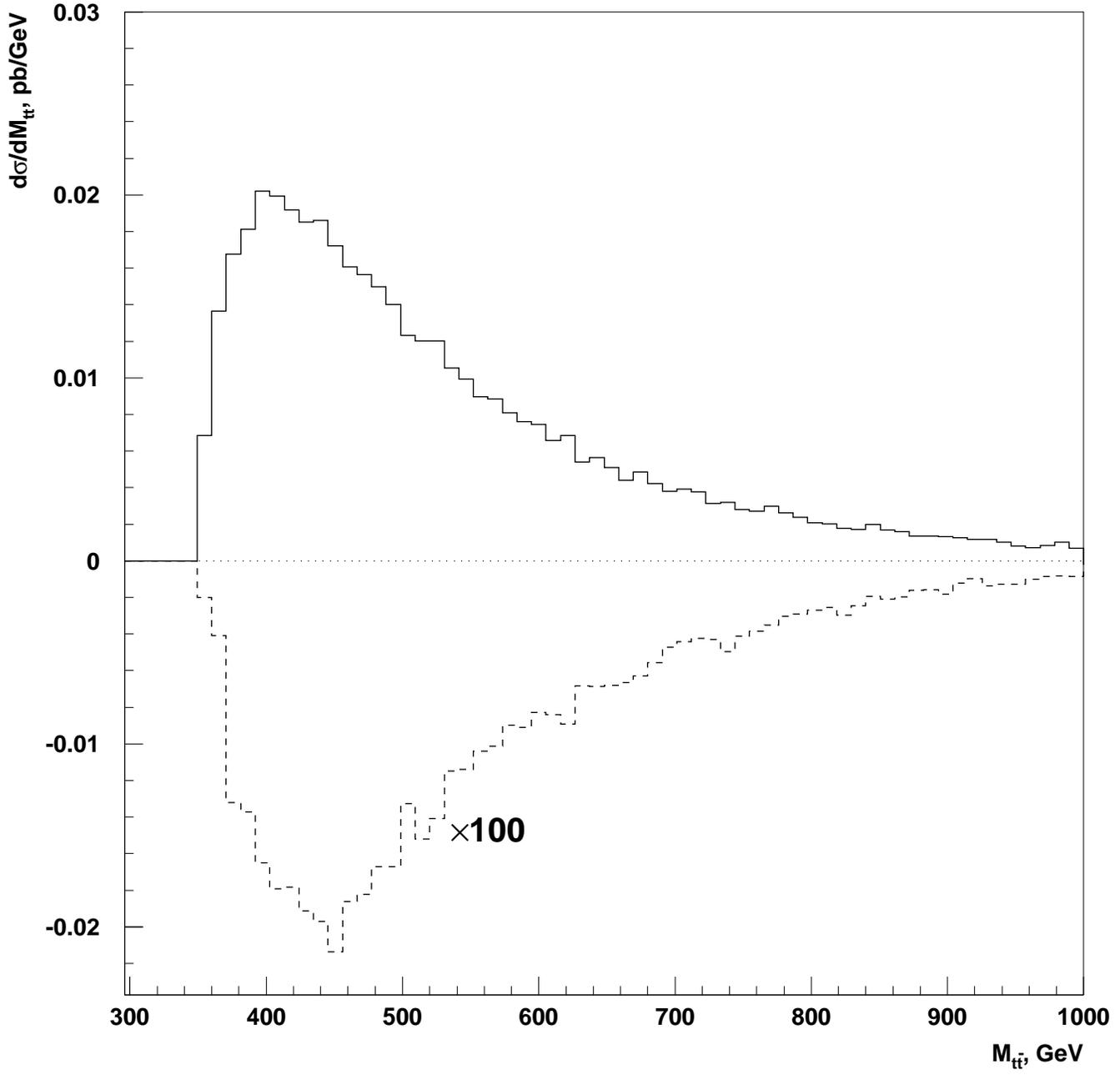
{\bf Fig. 3.} Dependence of the sum  
 $d\sigma_L/dM_{t\bar t}+d\sigma_R/dM_{t\bar t}$(solid line) and the 
difference $d\sigma_L/dM_{t\bar t}-d\sigma_R/dM_{t\bar t}$ 
 (dashed line, magnified by 100) of the differential
cross-sections of the production of the left- and right-handed tops
on the invariant mass of 
the $t\bar t$ pairs $M_{t\bar t}$. The factorization scale $\mu =2 m_t$,
the squark masses are (90,1000,0). 
\vspace{\baselineskip}

\noindent
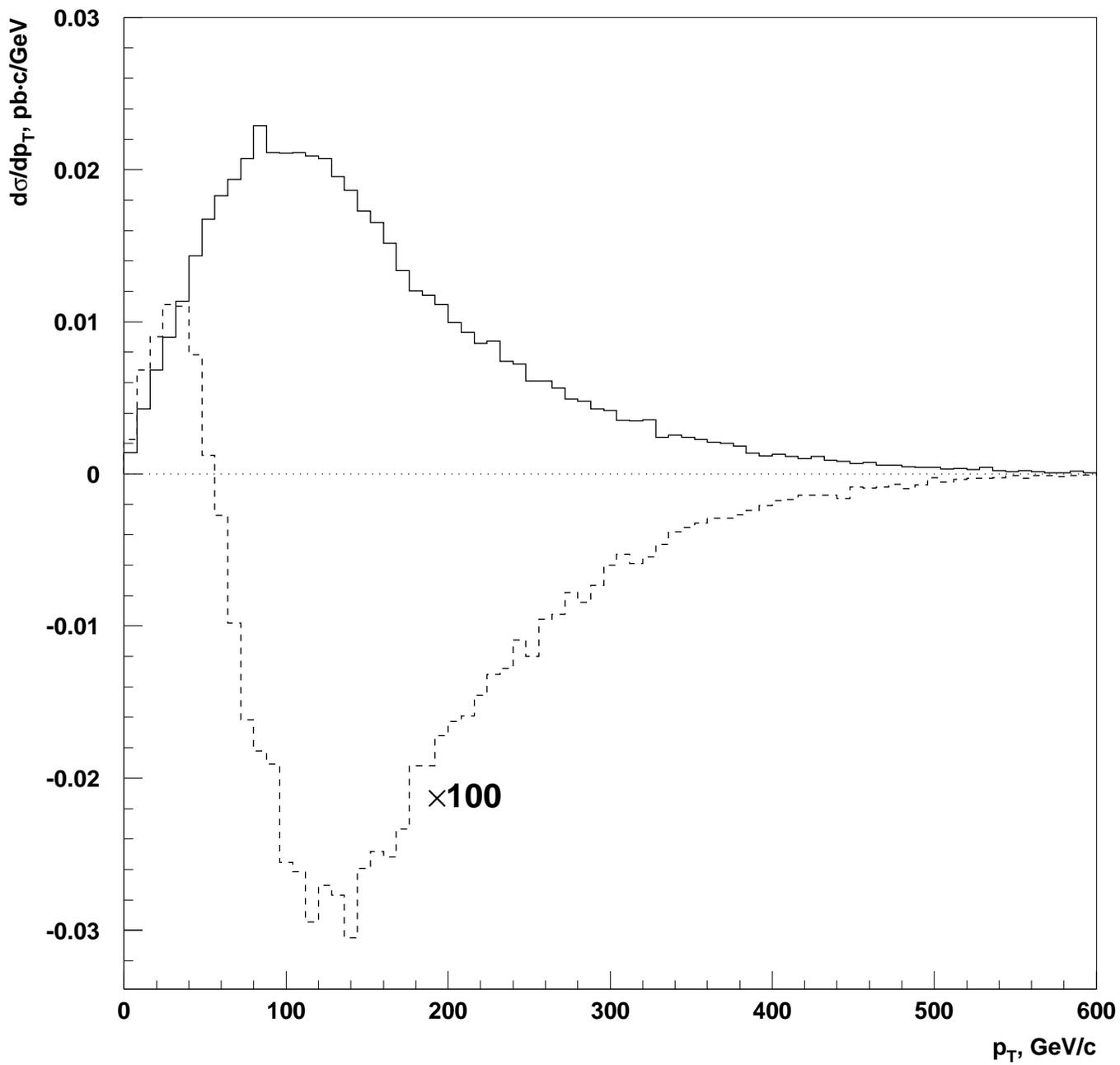
{\bf Fig. 4.} Dependence of the sum  
 $d\sigma_L/dp_{T}+d\sigma_R/dp_{T}$(solid line) and the 
difference $d\sigma_L/dp_{T}-d\sigma_R/dp_{T}$ 
(dashed line, magnified by 100)
of the differential
cross-sections of the production of the left- and right-handed tops
on the transverse momentum of 
the $t$-quark $p_{T}$. 
The factorization scale $\mu =2 m_t$, the squark masses are (90,1000,0). 
\vspace{\baselineskip}

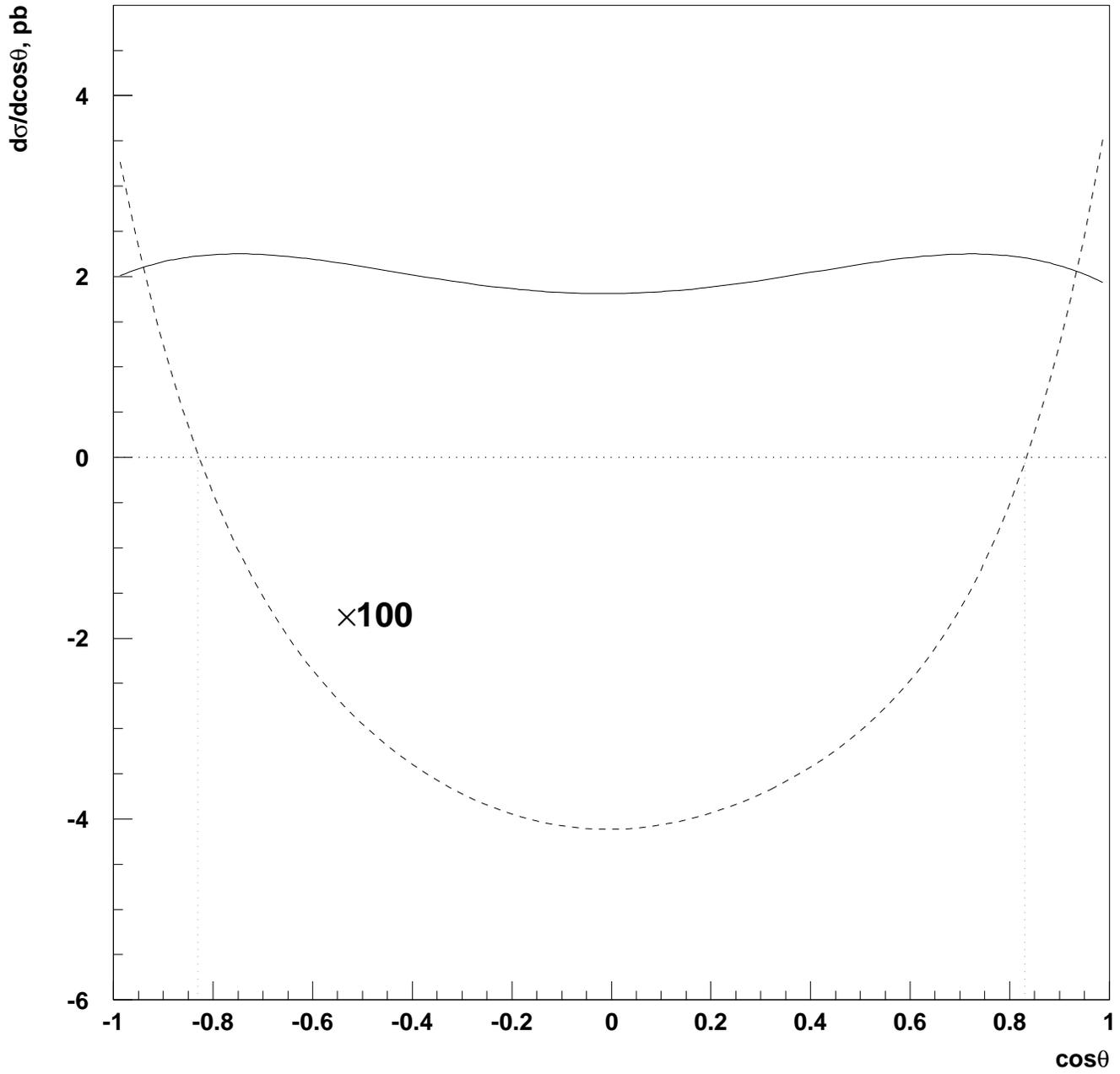
{\bf Fig. 5.} 
Dependence of the sum  
 $d\sigma_L/d\cos\theta+d\sigma_R/d\cos\theta$(solid line) and 
the difference $d\sigma_L/d\cos\theta-d\sigma_R/d\cos\theta$ 
(dashed line, magnified by 100)
of the differential
cross-sections of the production of the left- and right-handed tops
 on the cosine of the scattering angle in the 
 $t\bar t$ pair rest frame. 
The factorization scale $\mu =2 m_t$, the squark masses are (90,1000,0). 
\vspace{\baselineskip}

\newpage

\begin{table}[h]
\small
\begin{tabular}{||c||c|c||c|c||c|c||c|c||}
\multicolumn{1}{||c||}{}& \multicolumn{4}{c||}{    $\mu=m_t$} & 
 \multicolumn{4}{c||}{ $\mu=2 m_t$} \\
\cline{2-9}
\multicolumn{1}{||c||}{Masses (GeV)}& 
\multicolumn{2}{c||}{$\small |\cos\theta|\leq 0.8$} 
& \multicolumn{2}{c||}{$\small |\cos\theta|> 0.8$} 
&\multicolumn{2}{c||}{$\small |\cos\theta|\leq 0.8$} 
& \multicolumn{2}{c||}{$\small |\cos\theta|> 0.8$}  
\\
\cline{2-9}
($m_{{\tilde t}_1},m_{{\tilde t}_2},m_{LR}$)& $A_t$ &$\delta A_t$ &
$A_t$ &$\delta A_t$ & $A_t$ &$\delta A_t$ & $A_t$ &$\delta A_t$\\
\hline
\hline
  (90,150,0)  &  -0.31& 0.23  &  1.14& 0.54  &  -0.35& 0.25  &  1.20& 0.60  \\
\hline
  (90,200,0)  &  -0.57& 0.23  &  1.30& 0.54  &  -0.62& 0.25  &  1.42& 0.60  \\
\hline
  (90,500,0)  &  -1.31& 0.23  &  1.20& 0.54  &  -1.39& 0.25  &  1.33 & 0.60  \\
\hline
  (90,1000,0)  &  -1.50& 0.23  &  1.05& 0.54  &  -1.61& 0.25  &  1.20& 0.60  \\
\hline
  (175,250,0)  &  -0.33& 0.13  & --- & --- &  -0.36 & 0.14  & --- &---  \\
\hline
  (175,500,0)  &  -0.86& 0.13  & --- & ---  &  -0.91 & 0.14 & --- & ---  \\
\hline
  (175,1000,0)  &  -1.04& 0.13  & --- & --- &  -1.11 & 0.14 & --- & ---  \\
\end{tabular}
\vspace{\baselineskip}
\caption{The asymmetries $A_t$ (in \% ) predicted by SUSY QCD, 
and the estimated statistical errors of their 
measurement $\delta A_t$ for the LHC luminosity ${\cal L}=100\ fb^{-1}/year$.
The entries with the hyphen correspond to the asymmetries which are too small 
to be observed.}
\end{table}

\end{document}